\documentclass[%
 reprint,
]{revtex4-2}

\usepackage{graphicx}% Include figure files
\usepackage{dcolumn}% Align table columns on decimal point
\usepackage{bm}% bold math
\usepackage{color,soul}
\begin{document}

\preprint{APS/123-QED}

\title{Ultracompact, dynamically controllable circularly polarized laser enabled by chiral metasurfaces}

\author{Ioannis Katsantonis$^{1,3}$}
\email{katsantonis@iesl.forth.gr} 
\author{Anna Tasolamprou$^3$}%
\author{Eleftherios Economou$^{1,4}$}%
\author{Thomas Koschny$^5$}%
\email{koschny@ameslab.gov}
\author{Maria Kafesaki$^{1,2,}$}%
\email{kafesaki@iesl.forth.gr}
\affiliation{
 $^1$Institute of Electronic Structure and Laser, Foundation for Research and Technology Hellas, 70013, Heraklion, Greece
}%
\affiliation{
 $^2$Department of Materials Science and Technology, University of Crete, 70013, Heraklion, Greece
}%

\affiliation{$^3$Department of Physics, National and Kapodistrian University of Athens, 15784, Athens, Greece
}%

 \affiliation{
$^4$Department of Physics, University of Crete, 70013, Heraklion, Greece.
}%

\affiliation{
$^5$Ames Laboratory and Department of Physics and Astronomy, Iowa State University, Ames, Iowa, 50011, USA}%

\date{\today}% It is always \today, today,

\begin{abstract}
%Circularly polarized (CP) emission lasers are attracting increasing attention in photonics and materials science. 
We demonstrate a simple, low-cost and ultracompact chiral resonant metasurface design, which, by strong local coupling to a quantum gain medium (quantum emitters), allows to implement an ultra-thin metasurface laser, capable of generating tunable circularly polarized coherent lasing output.  According to our detailed numerical investigations the lasing emission can be transformed  from linear to circular and switch from right- to left-handed circularly polarized (CP) not only by changing the metasurface chiral response but also by changing the polarization of a linearly polarized pump wave, providing thus dynamic lasing-polarization control. Given the increasing interest for CP laser emission, our chiral metasurface laser design proves to be a versatile yet straightforward strategy to generate strong and tailored CP emission laser, promising great potential for future applications in both photonics and materials science.
\end{abstract}

\maketitle

%\tableofcontents

%\section{\label{sec:level1}IN%TRODUCTION}
Polarization controllable and, even more, circular polarization (CP) lasers, featuring significant prospects in spectroscopic, sensing and display technologies, are a growing area in the field of light-matter interactions \cite{Coles2010676, Hubener2021438, Lindemann2019212, Shree202139, Genevet2017139,Chen2019,Hajji202119905,Movsesyan2023,Stanciu2007}. CP lasers can also serve as valuable tools for investigating and comprehending chiral-light-matter interactions, an issue playing a critical role in various scientific disciplines, such as chemistry \cite{Wenzel20187261}, biophysics \cite{Naaman202299} and quantum optics \cite{Thanopulos2004228,Lodahl2017473,Zu2019775,Sherson2006557}. Achieving CP lasing, in principle, requires a combination of a gain medium and a chiral response  \cite{Qu20218753}. One way is to use natural chiral molecules \cite{Jimenez20}  combined with a gain medium \cite{Krasnok2020628}. However, this strategy is inefficient due to the very weak chiral response of the molecules. Another approach is based on chiral nematic liquid crystals combined with active (gain) molecules \cite{Kopp19981707,Fuh20041857,Xu202011130}. However, the requirement to synthesize the liquid crystal with the dyes or quantum dots or wells (as active "molecules") complicates the preparation process of the chiral nematic liquid crystal-based laser system.

Recently, chiral light sources based on two dimensional transition metal dichalcogenides (TMDCs) \cite{Zhang2014725,Liu2019,Yang201921367,herrmann2023valley} and perovskites \cite{Long2020423,Long2018528,kim2023chiral,Wang2019,Seo202113781} combined with metamaterials have attracted remarkable attention, due to their ability to give chiral photoluminescence. However, the degree of circular polarization (i.e. the polarization ellipticity of the emitted wave) demonstrated in such media remains weak, thus limiting their practical applications potential. 

An alternative of chiral "molecules" approach to achieve CP lasing is to employ chiral nanophotonic structures (operating as optical cavities). In recent years various nanophotonics structures have been proposed to modulate the environment of the quantum emitters and to enhance the radiation emission or/and control its polarization state \cite{Konishi2011,Ren20152951,Cotrufo20163389,Dorrah2022,Maksimov2022,Li20212893, Tang2011333,Sun2023,Carter2013329,Droulias2022}. Among these, chiral metasurfaces based on quasi-bound states in the continuum \cite{Zhang20221215} are remarkable due to the fact that they can exhibit a highly CP laser output. However, these chiral metasurfaces are not offered for dynamic polarization control and are not rigorously planar, which may cause cumbersomeness in their fabrication process. Therefore, a flexible strategy to develop controllable CP laser at large scale with low-cost and feasible fabrication remains challenging. Here we propose and investigate an approach based on a compact bi-layer-metal chiral metasurface with gain, and demonstrate its high potential and capabilities for controllable CP lasing. 

Chiral metamaterials (CMMs), i.e. artificial structures composed of building blocks with no mirror symmetry plane \cite{Plum2009,Decker20101593,Valev20132517,Hentschel2017,Oh2015,Gansel20091513}, are characterized by strong magneto-electric coupling, i.e. strong chirality, since it is not restricted by the atomic size as in natural chiral materials. Indeed, strong circular dichroism (i.e. absorption difference between left- and right-handed CP waves) and large polarization rotation (effects connected with the strength of chirality) have been demonstrated using chiral metamaterials \cite{Gansel20091513,Yin20136238,Katsantonis2023}, in various frequency ranges. These effects and possibilities make chiral metamaterials suitable for applications based on wave polarization control, such as ultrathin circular polarizers, polarization modulators, etc. Interestingly, it has been recently shown that combining chiral media with Parity-Time-symmetry \cite{Droulias2019,Katsantonis2020,Katsantonis2020Sc} in a simple double-layer system opens the way to a plethora of electromagnetic wave control applications, such as  circular polarization lasing, polarization isolation, as well as asymmetric (side-dependent) electromagnetic wave transmission \cite{Katsantonis2020}.

\begin{figure}[ht!]
\includegraphics[width=2.8in]{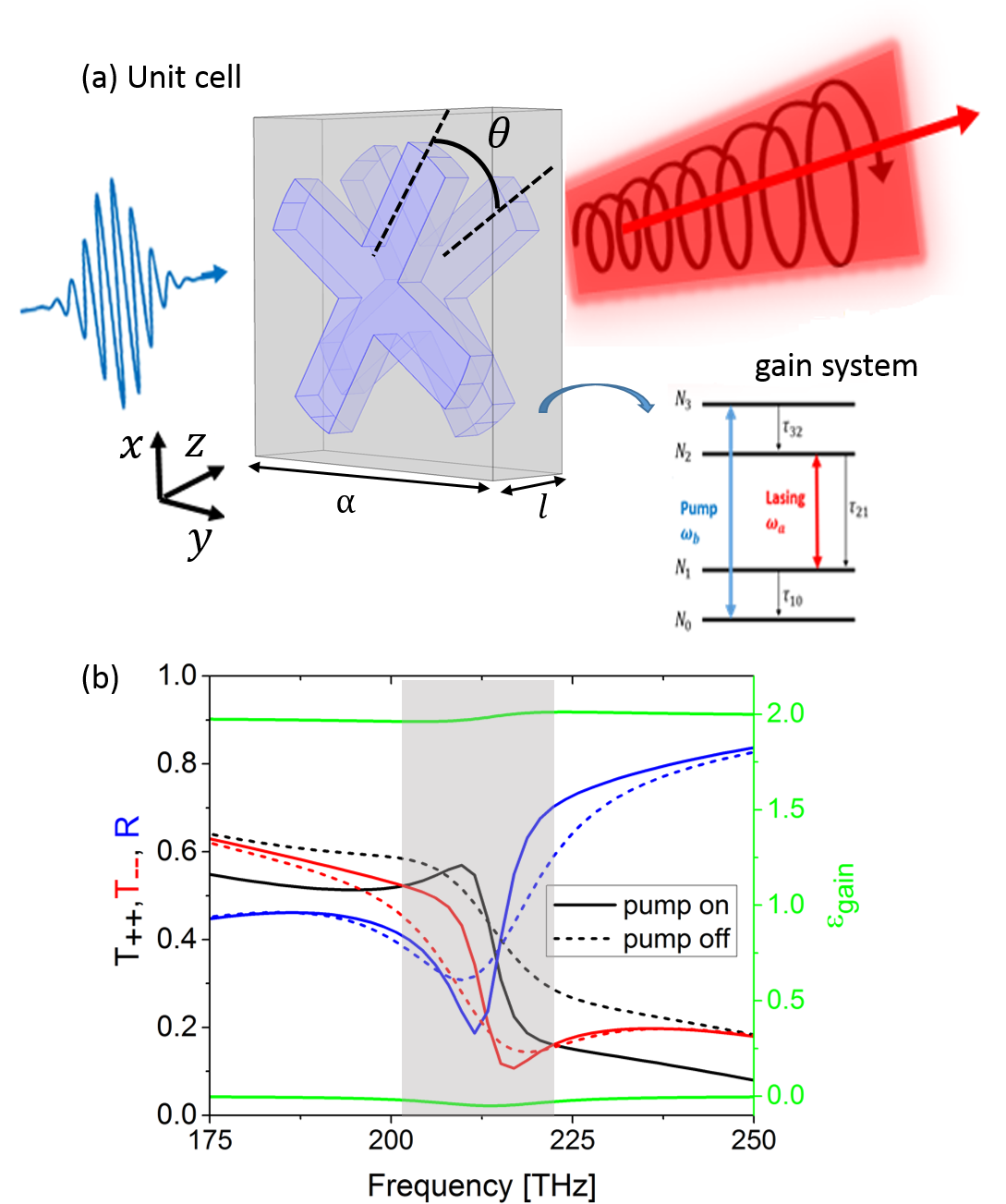}% Here is how to import EPS art
\caption{\label{fig1}(a) Schematic illustration of the unit cell for the silver-based twisted-crosses chiral metasurface and the employed four level gain system. (b) Calculated linear spectra for transmittances and reflectances for the structure shown in Fig.\@ 1 (a). The green lines show the profile of the gain medium [real (solid line) and imaginary (dashed line) part of its electric permittivity - see Eq. (6)] for pump rate \(R_{p}=1.5 \times 10^{8}\) s\(^{-1}\). The shadow highlights the gain material bandwidth.}
\end{figure}

Inspired by these developments, here we introduce the concept of a resonant bi-isotropic chiral metasurface, locally strongly coupled to a thin, realistic quantum gain medium, as a promising approach to ultracompact, controllable CP wave metasurface lasers. The linear scattered field of our chiral metasurface arises from the collective radiated field of the periodic resonant chiral meta-atoms. Properly designed, it can be made purely circular polarized. Strong coupling to the quantum gain material mediated by the resonant near-field of the chiral meta-atoms will compensate the losses in these plasmonic resonators and eventually have them spontaneously oscillate coherently, i.e., drive them into laser state. The lasing output of this surface laser is the circularly polarized radiated field of the lasing resonant plasmonic eigenmode of the chiral metasurface, where the chiral crossed-wires resonator effectively constitutes the sub-wavelength, plasmonic "resonant cavity" of the laser, allowing for strongly enhanced light-matter interaction and sub-wavelength size. For the chiral metasurface we employ a design based on two mutually twisted metallic  crosses (as, e.g., in Ref.\cite{Zhang2009}) and validate the polarization controllable lasing by numerical simulations. 
Adjusting the twist-angle between the crosses or, more importantly, adjusting dynamically the input polarization angle enables the manipulation of the chiral response, which couples with the optical gain band, thus allowing different laser emission behaviors to be demonstrated.  Our active (gain) chiral metasurface provides an opportunity to design lasing with any desired polarization state, including CP, eliminating the need for expensive and tedious fabrication.

\begin{figure}[ht!]
\includegraphics[width=3.4in]{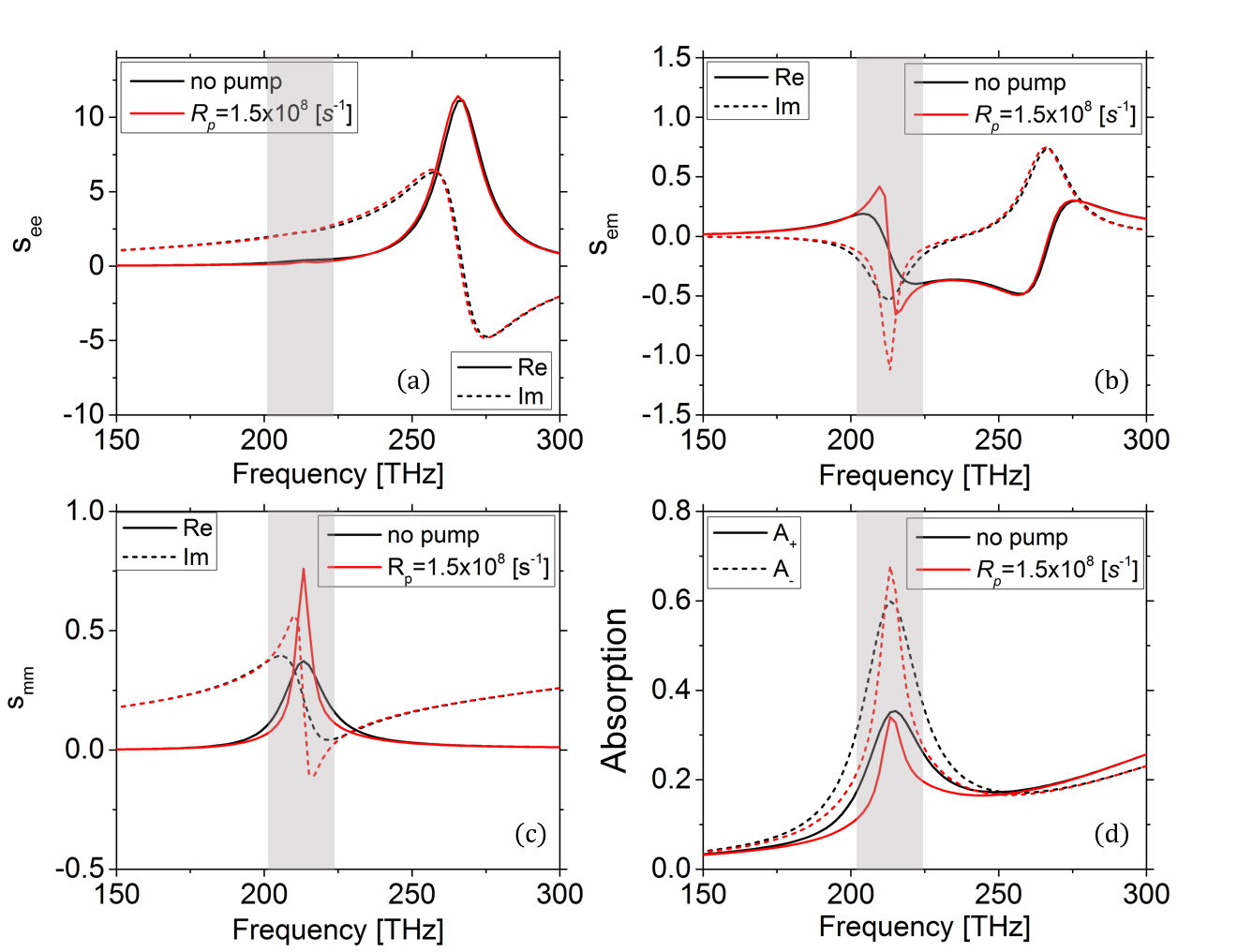}% Here is how to import EPS art
\caption{\label{fig2} Real (solid lines) and imaginary (dashed lines) parts of the effective sheet conductivities (dimensionless quantities), i.e. the electric sheet conductivity (a), the magnetic (b), and the magneto-electric (c), for the system of Fig.\@ 1, for pump rates \(R_{p}=0.0\,\mathrm{s}^{-1}\) and  \(R_{p}=1.5 \times 10^{8}\,\mathrm{s}^{-1}\). Panel (d) shows the absorption amplitudes for RCP/+ and LCP/- waves for the same system and pump-rates. The shadow regime corresponds to the gain resonant response for pump rate \(R_{p}=1.5 \times 10^{8}\,\mathrm{s}^{-1}\).}
\end{figure}

The unit cell of the chiral metasurface considered here is shown in Fig.\@ \ref{fig1}(a). It consists of two mutually twisted metallic crosses, of twist angle \(\theta=22.5\)\(^{\circ}\) (the first cross is rotated \(11.25^{\circ}\) to the left with respect to the diagonal while the second \(11.25^{\circ}\) to the right), embedded in a dielectric host of refractive index \(n=1.41\). The structure (metasurface) has square periodicity (along $x,y$ directions) with lattice constant \(a=300\) nm, and thickness (along $z$-direction) \(l=100 \) nm. The metallic crosses are made of silver, described here by a Drude model: \(\epsilon(\omega)=\epsilon_{\infty}-\omega^{2}_{p}/(\omega^{2}+i\omega\gamma)\) with \(\epsilon_{\infty}=9.07\), \(\omega_{p}=2\pi\times2159 \times10^{12} \) rad/s and  \(\gamma=2\pi \times25 \times10^{12}\) rad/s. The additional geometric parameters of the metallic crosses are their height (side-length), \(h=250 \) nm, the width, \(w=56\) nm and the metal-thickness (along $z$), \(s=25\) nm. Figure 1 (b) illustrates the calculated linear spectrum (with and without gain) of transmittances \(T_{++}\) and \(T_{--}\) and reflectances \(R=R_{+-}=R_{-+}\) for the structure of Fig.\@ 1 (a), for circularly polarized incident wave, where the first (second) subscript denotes the output (input) wave polarization, and + and - indicate right-handed circularly polarized (RCP) and left-handed circularly polarized (LCP) wave respectively. Note that the cross-polarized transmittances and co-polarized reflectances are zero (see Supplemental Material). Fig.\@ 1 (b) shows a resonance at frequency around 214 THz, which is a predominantly magnetic resonance rendered chiral by the twist angle between the crosses (with antiparallel currents at the two "facing" crosses - see Supplemental Material; also at \cite{Wang2009}). This local magnetic resonance is strongly coupled to the structure-material with the near field. Such a coupling in the presence of gain leads to reshaping and un-damping of the spectral response as depicted in Fig.\@ \ref{fig1}(b) and explained in more detail below (Fig.\@ \ref{fig2}). 
 
The gain material can be obtained by doping the dielectric host with dyes, described here as 4-level systems, as depicted in Fig.\@ 1 \cite{Fang2010,Hess2012573,Wuestner2010,Droulias2017131,Droulias2017}. The population density in each level is given by \(N_{i}\)  \((i=0,1,2,3)\). We assume that initially all the electrons are in the ground state \((N_{0})\). Next, the electrons are pumped by an external electromagnetic wave with frequency \(\hbar\omega_{b}=E_{3}-E_{0}\) where \(E_{3}\), \(E_{0}\) are the energies of the excited (third level) and ground state (zero level), respectively. After a short lifetime \(\tau_{32}\), the electrons relax into the metastable level,  level 2. By spontaneous and stimulated emission as well as by non-radiative processes the electrons are transferred to the first level, level 1.  Levels 1 and 2 are  the lasing states and the lasing frequency is \(\hbar\omega_{a}= E_{2}- E_{1}\). The lifetimes and energies of the lasing levels are \(\tau_{21}\), \(E_{2}\) and \(\tau_{10}\), \(E_{1}\). The atomic populations at each spatial point obey the following rate equations \cite{Fang2010,Hess2012573,Wuestner2010,Droulias2017131,Droulias2017}: 
\begin{eqnarray} 
\label{eq:1}
\dot N_{3} &=& (\hbar\omega_{b})^{-1}\textbf{E} \cdot \dot\textbf{P}_{G}^{(b)} -(\tau_{32})^{-1}N_{3} \\
\label{eq:2}
\dot N_{2} &=& (\tau_{32})^{-1}N_{3}+ (\hbar\omega_{a})^{-1}\textbf{E} \cdot \dot\textbf{P}_{G}^{(a)} -(\tau_{21})^{-1}N_{2} \qquad \\
\label{eq:3}
\dot N_{1} &=& (\tau_{21})^{-1}N_{2}- (\hbar\omega_{a})^{-1}\textbf{E}  \cdot\dot\textbf{P}_{G}^{(a)} -(\tau_{10})^{-1}N_{1} \\
\label{eq:4}
\dot N_{0} &=& -(\hbar\omega_{b})^{-1} \textbf{E} \cdot \dot \textbf{P}_{G}^{(b)} + (\tau_{10})^{-1} N_{1}
\end{eqnarray}
where \(\textbf{P}_{G}^{(a)}\) is the induced electric polarization density of the atomic transition between the lasing levels (1 and 2, radiation emission), \(\textbf{P}_{G}^{(b)}\) is the induced electric polarization density of the transition between ground and excited state (0 and 3, pumping, radiation absorption) while the dots on the populations and polarizations indicate time derivative. The induced macroscopic polarization is related with the microscopic polarization of the molecules (see Supplemental Material) and is coupled to the local electric field \(\textbf{E}\)  by the following equation: 
\begin{equation} \label{eq:5}
 \ddot\textbf{P}_{G}^{(a,b)}+\Gamma_{a,b} \dot \textbf{P}_{G}^{(a,b)} + \omega_{a,b}^2 \textbf{P}_{G}^{(a,b)}=-\sigma_{a,b} \Delta N_{a,b} \textbf{E}
\end{equation}
%\begin{equation} %\label{eq:5}
%\frac{\partial^2 %\textbf{P}_{G}^{(a,b)}}{\partial t^2}+\Gamma_{a,(b)} \frac{\partial \textbf{P}_{G}^{(a,b)}}{\partial t} + \omega_{a,(b)}^2 \textbf{P}_{G}^{(a,b)}=-\sigma_{a,(b)} \Delta N_{a} %\textbf{E}
%\end{equation}
where \(\Delta N_{a}=N_{2}-N_{1}\), \(\Delta N_{b}=N_{3}-N_{0}\), \(\Gamma_{a}\), \(\Gamma_{b}\) are the linewidths of the atomic transitions at the emitting angular frequencies and \(\sigma_{a}\), \(\sigma_{b}\) are the coupling strength of the \(\textbf{P}_{G}^{(a)}\), \(\textbf{P}_{G}^{(b)}\) to the electric field \(\textbf{E}\) whose values can be obtained experimentally.
Within this framework, the Maxwell's equations are coupled with the rate equations (\ref{eq:1})-(\ref{eq:4}) via Eqn.\@ (\ref{eq:5}), through the polarization density, while all equations share the same simulation domain, along with the same spatial and temporal discretization. Here we integrate our four-level formalism into the chiral metasurface and perform the combined system analysis using Finite Element Time Domain (FETD) simulations through the COMSOL Multiphysics software. The parameters for the four-level system are chosen as follows \cite{Fang2010,Droulias2018}: 
Total electron density \(N_{el} = N_{0}(t=0) = 5\times 10^{23}\)  m\(^{-3}\), coupling coefficients   \(\sigma_{a}=10^{-4}\) C\(^{2}/\)kg, \(\sigma_{b}=5\times 10^{-6}\)  C\(^{2}\)/kg,  linewidths \(\Gamma_{a}=2\pi \times 10 \times 10^{12}\) rad/s, \(\Gamma_{b}=2\pi \times 20 \times 10^{12}\) rad/s and relaxation times \(\tau_{10}=\tau_{32}=0.05\) ps, \(\tau_{21}=20\) ps.

% Thomas - begin changes in progress

In order to better understand the population dynamics and the effective polarizablility (permittivity) contributed by the gain system,
we consider the stationary state lasing, replacing the pump field by an average pumping rate, approximating the rate of absorbed pump photons per volume,
\( (\hbar\omega_{b})^{-1}{\bf E} \cdot {\dot{\bf P}}_{G}^{(b)} \), by an abstract pumping rate of electrons from the ground level to the top level, 
\(N_0 R_p\).
The pump rate \(R_{p}\) is related to the absorbed pump intensity, $I_{pump}$, via \(N_0 R_{p} = I_{pump}/(\hbar\omega_{a}\, l)\), 
where \(l\) is the thickness of the gain layer within the metasurface. 
Then the macroscopic polarization density induced by the gain medium can be incorporated in the frequency-dependent constitutive relations:
\(\textbf{P}_{G}^{(a)} = \big(\varepsilon_g(\omega)-\varepsilon_h\big)\, \varepsilon_0 \textbf{E}\), where \(\varepsilon_{h}\) is the host relative  permittivity.
Applying the steady state approximation assuming weak fields 
in Eqn.\@  (\ref{eq:1})-(\ref{eq:4}) and (\ref{eq:5}), 
neglecting all nonlinear terms except the leading order 
(which is constant in the population numbers, \( dN_{i}/dt \sim 0,\ i \in \{0,1,2,3\}) \), and linear harmonic time-dependence in the electric field and polarizations,
we can express the relative permittivity of the gain-host material as \cite{Caligiuri20171012,Coppolaro20202578,Droulias2018}:
\begin{equation} \label{eq:7}
\varepsilon_{g}(\omega) \ =\  
  \varepsilon_{h} \ + \frac{1}{\varepsilon_0}\sum\limits_{m\in\{a,b\}} \frac{-\sigma_{m}\, \Delta N_{m}}{\omega_{m}^{2}-\omega^{2}-i\omega\, \Gamma_{m}}
\end{equation}
where
\( \Delta N_a = (\tau_{21}-\tau_{10}) R_p\, N_0 \),
\( \Delta N_b = (\tau_{32} R_p -1) N_0 \), and
\( N_0 = [ 1+(\tau_{32}+\tau_{21}+\tau_{10}) R_{p} ]^{-1} N_{el} \).
From the last equation, under the chosen conditions, we find that the contribution from the lasing transition is about one order of magnitude smaller than the host permittivity \(\varepsilon_h\) which will affect the metasurface by renormalzing the resonances of the chiral meta-atoms, both in frequency and damping,
while the correction from the pump transition is yet another order of magnitude weaker and mostly negligible except for the fact that it renders the lasing resonance of the chiral metasurface slightly sensitive to the intensity (and polarization) of the pump radiation.
For the metasurface shown in Fig.\@ 1(a) and pump rate $1.5\times 10^8\,\mathrm{s}^{-1}$ this permittivity \(\varepsilon_g(\omega)\) is shown in Fig.\@ 1(b). Fig.\@ 1(b) also shows the metamaterial response for pump on, i.e. for the host of relative permittivity as in Eq. \ref{eq:7}.  Note that the coupling of the chiral structure with the gain material results to un-damping of the metamaterial resonance, as is expected (see also \cite{Droulias2019N}). This change is highly affected by the pump rate, $R_p$.

% Thomas - end changes in progress

To quantify the chiral structure response and examine how it is affected by the gain we use the standard retrieval procedure \cite{Droulias2020} suitable for thin metamaterial layers/sheets, to extract the effective dimensionless sheet conductivities, \(s_{ee}\), \(s_{mm}\), \(s_{em}\), of the structure with and without gain (see also effective material parameters in the Supplemental Material). Figure \ref{fig2}(a) shows the retrieved results for the real (solid lines) and the imaginary (dashed lines) parts of the effective electric conductivity \(s_{ee}\), with gain (different pump rates) and without gain. One can see that increasing the pump rate, and hence the available gain, leads to a slight decrease in the real part of the electric dimensionless conductivity in the frequency range close to the maximum emission cross-section of the gain medium, demonstrating reduced losses (in agreement with Fig.\@ 1(b)).  Figure \ref{fig2}(b) shows the real (solid lines) and imaginary (dash lines) parts of the effective magnetic conductivity \(s_{mm}\), with and without gain. We observe that with the incorporation of gain, the weak and broad magnetic resonance of the passive metasurface becomes strong and narrower, indicating that that the electric gain counteracts the magnetic losses. 
Besides effective electric and magnetic conductivities, we also calculate the effective sheet magneto-electric conductivity \(s_{em}\), a measure of the chiral response of the structure. Figure \ref{fig2}(c) illustrates the real and the imaginary parts of  \(s_{em}\), with and without gain. We see that increasing the pump rate \begin{figure}[ht!]
\includegraphics[width=3.4in]{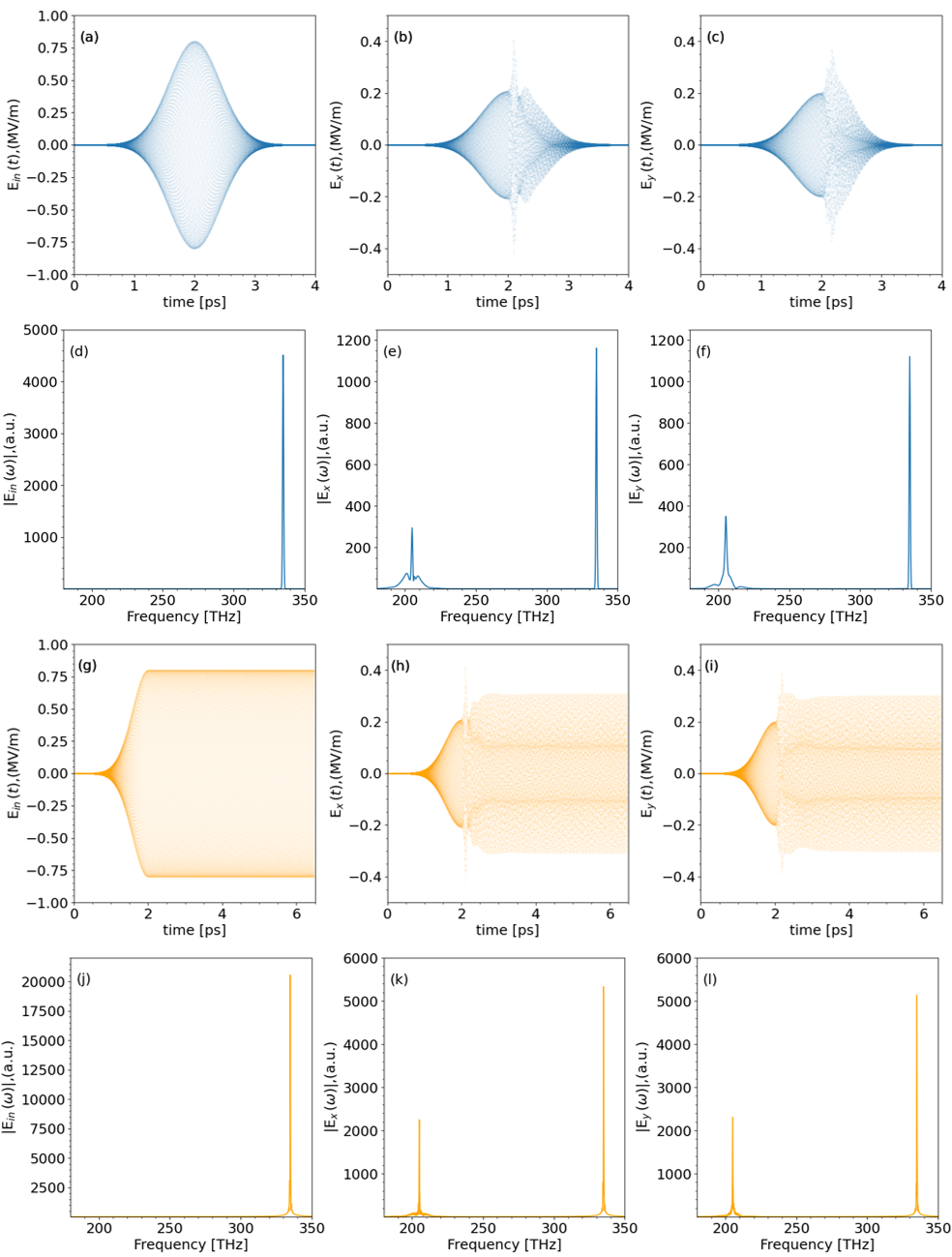}% Here is how to import EPS art
\caption{\label{fig3}
The incident and transmitted waves as a function of time for the structure of Fig.\@ 1 excited by a pump pulse [panels (a)-(c)] and continue waves (CW) [panels (g)-(i)] of frequency 335 THz and polarization angle $\phi=45^{\circ}$. Panels (a), (b) and (c) are the incident, transmitted \(E_{x}\), and transmitted  \(E_{y}\) waves for input pump-pulse amplitude \(0.8 \times 10^{6}\) V/m. Panels (d), (e) and (f) are the corresponding Fourier transformed spectra of the incident and transmitted waves. Panels (g), (h) and (i) are the incident, transmitted \(E_{x}\), and transmitted  \(E_{y}\) waves, as a function of time, for input/pump CW amplitude \(0.8 \times 10^{6}\) V/m. Panels (j), (k) and (l) are the corresponding Fourier transformed spectra. The sharp peaks at about 335 THz is the incident beam (driving the gain from level 0 to 3).}
\end{figure} leads to a strong enhancement of both real and imaginary parts of \(s_{em}\), i.e. a strong enhancement of the system chiral response (see Supplemental Material for details). 

In order to clarify and quantify the impact of gain on the polarization state of the transmitted wave through the system, we calculate the absorption for RCP (+) and LCP (-) waves (determining the wave ellipticity; for the corresponding ellipticity see Supplemental Material) - see Fig.\@ 2(d). One can see that the increasing of pump rate reduces the absorption peak for LCP waves, while it narrows the peak for RCP waves without affecting much its maximum value. The corresponding ellipticity, which without gain is close to 20\(^{\circ}\), for \(R_{p}=1.5 \times 10^{8}\) s\(^{-1}\) gets close to 40\(^{\circ}\), indicating pure circularly polarized waves.  Hence, we find that the  effective amplification in the gain-metasurface system, which is controllable via pump rate, highly affects the polarization state of the output wave, allowing dynamic polarization tuning. It is worth-noting here that the LCP absorption ($A_-$) is more sensitive to the gain value; as we will show later, it is this mode (LCP) which will first lead to lasing. 

% we will say it later mode as clearly shown comparing Fig.\@ \ref{fig2} (d) with Fig.\@ \ref{fig5} (c). This mechanism can be understood by the asymmetric geometry along the propagation direction, the transmission responses for \(T_{--}\) and \(T_{++}\)  split into two curves which each one couples with gain resonance leading to the enhancement of the asymmetry.

In the above analysis we show the strong coupling between gain and chiral response, resulting to gain-controllable transmitted wave ellipticity; however, the main objective of this study is to demonstrate controllable (including circular) polarization lasing in our structure. For this demonstration and the understanding of the underlying physical mechanisms, we examine the active chiral metasurface of Fig.\@ \ref{fig1} (a) employing the "quantum" approach for the gain, i.e. Eqn.\@ (1)-(5), and linearly polarized input/pump wave. We first pump the gain molecules with a short intensive Gaussian pump pulse with a central frequency of \(\omega_{b} = 2\pi \times335\times10^{12} \) rad/s and duration \(t_{p} = 2\times\texttt{FWHM} = 2 \mathrm{ps}\), where \texttt{FWHM} denotes the full-width at half maximum of the pulse; 
\(\textbf{E}_p(t)=(\hat{\textbf{x}}\,\cos{\phi} +\hat{\textbf{y}}\,\sin{\phi} ) E_{p} \sin{(\omega_{b}\tau_{p})} \exp{[-\frac{1}{2}(\tau_{p} / \sigma_{p})^{2}]}\), 
where \(\tau_{p}=t-t_{p}\) and \(\sigma_{p}=\texttt{FWHM}/(2\sqrt{2\log2})\). Figures \ref{fig3}(a)-(c) depict the incident, co-polarized transmitted and cross-polarized transmitted waves as a function of time for polarization angle \(\phi=45\)\(^{\circ}\). Then, we Fourier transform the time-dependent transmitted electric fields to see if there is emission and how strong is the emitted radiation around \(\omega_{a}=2\pi \times 205 \times 10^{12}\) rad/s (lasing frequency). The results are depicted in Figs.\@ 3(d)-(f). The pump-pulse amplitude, $E_p$, is \(0.8 \times 10^{6}\) V/m and we do observe a peak at the emission frequency. An even higher amplitude of the pump pulse leads to higher emitted/lasing power. A worth-noticing feature in Fig.\@ 3 is the equal amplitudes of the $x$ and $y$ components of the lasing fields (see (e), (f)), indicating the possibility of circularly polarized emitted wave. Observing the relative phase between the two components we see that it is close to $\pi/2$, verifying the circular polarization character of the emitted wave.  

\begin{figure}[ht!]
\includegraphics[width=3.4in]{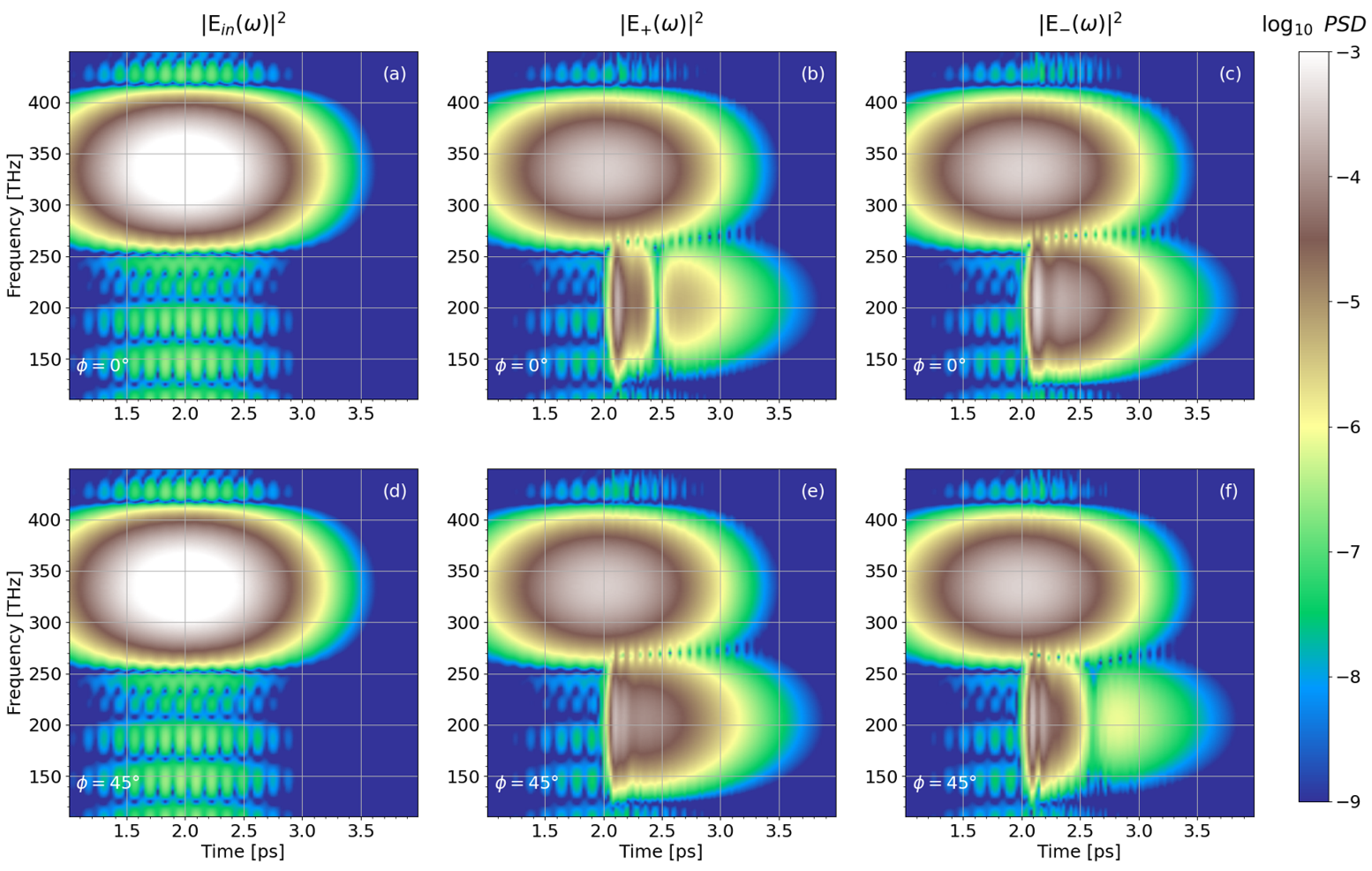}% Here is how to import EPS art
\caption{\label{fig4} The incident and transmitted power spectral density (color) for different polarization angles. Panels (a), (b) and (c) show the incident, RCP transmitted, \(|E_{+}|^2\), and LCP transmitted \(|E_{-}|^2\) wave, respectively, as a function of time and frequency for incident polarization angle \(\phi=0^{\circ}\). Panels (d), (e), (f) show the corresponding results for polarization angle \(\phi=45^{\circ}\).}
\end{figure}

To investigate further the lasing wave polarization and examine the possibility of its control, we keep the same pump-pulse amplitude, \(0.8 \times 10^{6}\) V/m,  we modify the relative polarization angle, $\phi$, from $\phi=0^{\circ}$, corresponding to linearly polarized incident pulse along \(x\)-axis, to $\phi=90^{\circ}$, corresponding to linearly polarized incident pulse along \(y\)-axis. The results for $\phi=0^{\circ}$ and $\phi=45^{\circ}$ are shown in Fig.\@ \ref{fig4}, where the two rows denote the two different angle cases, and both input and output waves are transformed in the circular polarization basis.  Figs.\@ \ref{fig4} (a), (d) show the power spectral density (in log scale) of the incident, (b), (e) of the RCP transmitted (\(E_{+}\)) and (c), (f) of the LCP transmitted (\(E_{-}\)) waves. When the polarization angle is $\phi=0^{\circ}$ we observe that above 2.5 ps the $E_{+}$ polarized emitted wave vanishes, while the $E_{-}$ polarized wave is the dominant, indicating the pure circularly polarized output. As we tune the polarization angle to \(\phi=45\)\(^{\circ}\) (the incident pulse is now polarized parallel to the diagonal of the metasurface unit-cell), we observe an interchange between the $E_{+}$ and $E_{-}$ transmitted waves at the lasing frequency; thus the dominant transmitted wave is the RCP, $E_{+}$, wave. (Further increase of the polarization angle, i.e. \(\phi=90\)\(^{\circ}\), leads again to LCP, $E_{-}$, transmitted output; intermediate cases are shown in the Supplemental material). This indicates that by modifying the angle $\phi$ one can dynamically tune the polarization state of the lasing mode, going from linear to circular polarization, and from RCP to LCP,  a capability of high importance in all applications based on or affected by the wave-polarization.
 
%\begin{widetext}

%\end{widetext}
\begin{figure}[ht!]
\includegraphics[width=3.4in]{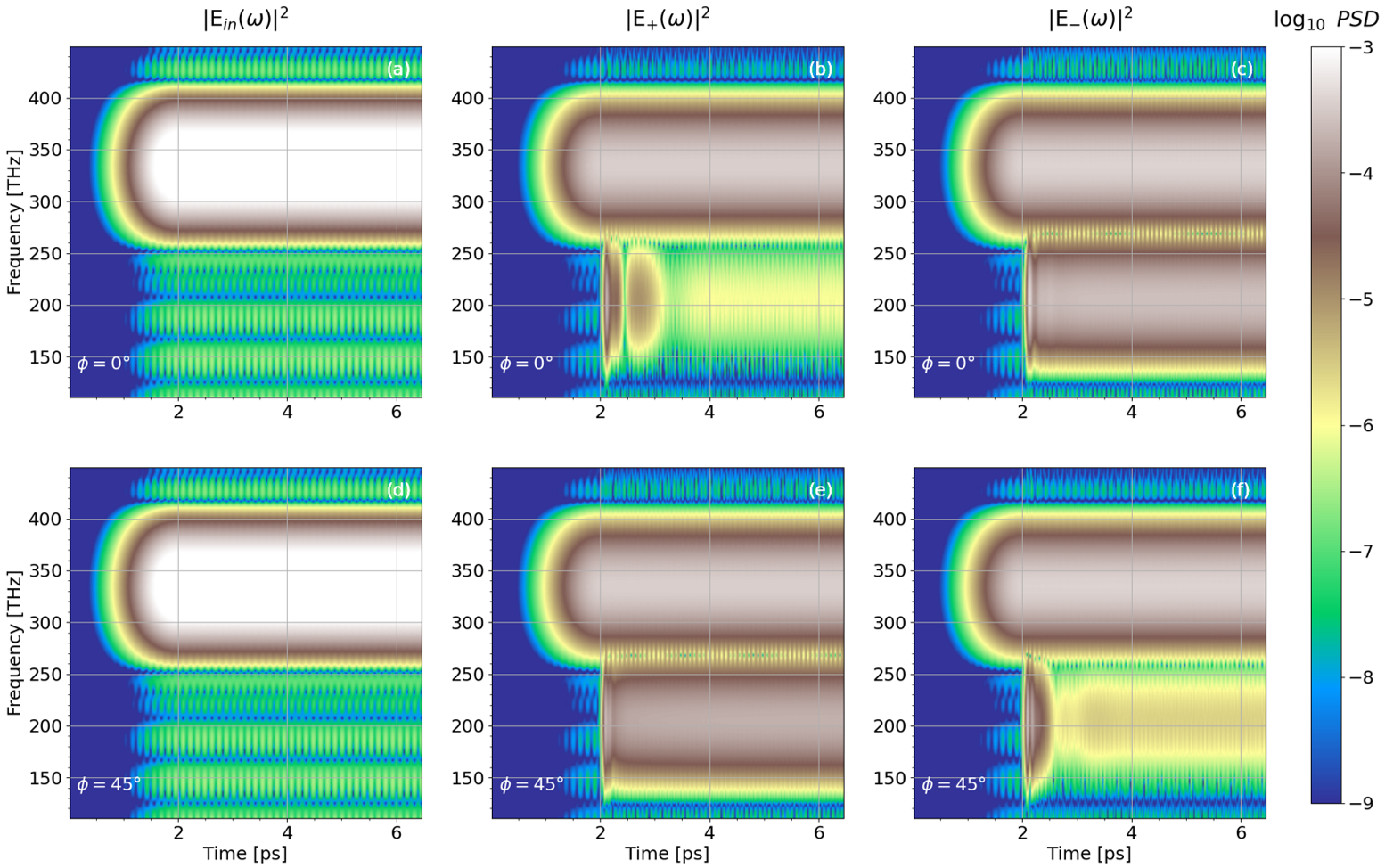}% Here is how to import EPS art
\caption{\label{fig5}
The incident and transmitted power spectral density (color) from the structure of Fig.\@ 1 for incidence of a continuous linearly polarized wave (CW) of different polarization angles, \(\phi\). Panels (a), (b) and (c) show the incident, RCP transmitted, \(|E_{+}|^2\), and LCP transmitted, \(|E_{-}|^2\), wave, respectively, as a function of time and frequency for incident polarization angle \(\phi=0^{\circ}\). Panels (d), (e), (f) show the corresponding results for incident pump polarization angle \(\phi=45^{\circ}\).}
\end{figure}

In the previous example we demonstrated that it can be achieved an input-polarization-controllable circular polarization laser but for a very short duration, as we excite the system with a pulse of relatively short duration. To examine further the impact of the input polarization angle on the output wave polarization, we calculated the  RCP/+ and LCP/- transmitted waves for our chiral metasurface under continuous wave (CW) excitation. The results are shown in Fig.\@ \ref{fig5}. Observing the results of Fig.\@ \ref{fig5}, we see that they follow closely the ones of Fig.\@ 4. I.e., for \(\phi=0^{\circ}\) the target lasing mode is dominated by LCP/-, with negligible RCP/+  emission. Twisting the polarization angle by \(\phi=45^{\circ}\) things reverse and the main lasing mode is dominated  by RCP/+ radiation (with LCP/-  emission  at the noise level). Corresponding results for intermediate angles demonstrate further the  input-polarization controllable laser output. 

\begin{figure}[ht!]
\includegraphics[width=3.4in]{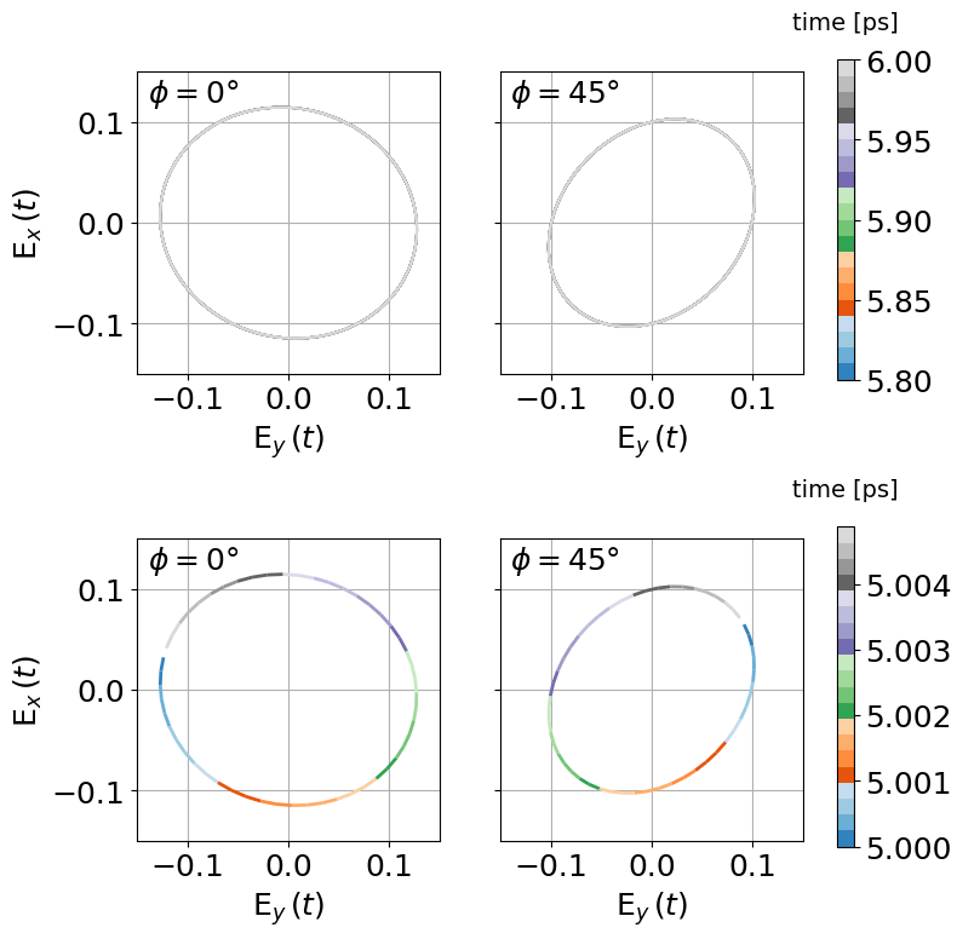}% Here is how to import EPS art
\caption{\label{fig6}
The transmitted waves \(E_{x}\)  versus \(E_{y}\)  for different polarization angles of the incident wave (\(\phi=0^{\circ}\) and \(\phi=45^{\circ}\) for a specific time duration, denoted in the colobar. Each panel is marked by the corresponding angle.}
\end{figure}

To confirm and investigate further the polarization of the outgoing laser radiation, we calculate also the linear components of the transmitted waves, \(E_{y}\) versus \(E_{x}\), for a specific time duration. The results are shown in Fig.\@ \ref{fig6}. We observe that at \(\phi=0^{\circ}\) an almost perfect circularly polarized emission laser is achieved. To determine also the handedness of the output laser we calculate the transmitted waves, \(E_{x}\) versus \(E_{y}\), over a single period. The results indicated by the color in \ref{fig6} clearly show that at \(\phi=0^{\circ}\) we observe an anti-clock wave while at \(\phi=45^{\circ}\) a clockwise wave. The above analysis clearly shows that our structure and approach can lead to controllable and pure circular polarization laser. To the best of our knowledge, up to date, there are no efficient routes to demonstrate and control pure (with a high degree of circular polarization) CP lasing in an ultra-compact form.

Closing, we would like to highlight here an at first look unexpected feature observed by comparing the results of Figs.\@ 2(d) and 5; we can observe that the lasing mode of Fig.\@ 5(c) (LCP mode) is the mode which  below lasing threshold is associated with the higher absorption and thus lower transmittance compare to the other CP mode. In other words, while our chiral structure in the absence of gain shows large cicrular dichroism resulting to highly RCP transmittance when it lases it first emits LCP wave. This is not surprising though, taking into account that emission is in fact inverse-absorption, and both processes are determined by the same material characteristics.  Thus, the mode of the higher absorption below lasing threshold is the one that first emits/lases above lasing threshold. 

In conclusion, we have presented a simple planar, low-cost, and ultracompact active chiral resonant metasurface design that allows easy fabrication of an ultra-thin metasurface laser capable of generating circularly polarized coherent lasing output. Coherent, circularly polarized output arises from direct lasing action of the collective resonant plasmonic eigenmode of the periodic, resonant chiral meta-atoms of the metasurface. Strong coupling to the quantum gain material mediated by the resonant near-field of the meta-atoms will compensate the dissipative losses in plasmonic twisted crosses-wire resonators and eventually have them spontaneously oscillate coherently, i.e., drive them into a lasing state. The lasing output of this surface laser is the circularly polarized radiated field of the lasing resonant plasmonic eigenmode of the chiral metasurface, where the chiral crossed-wires resonator effectively constitutes the sub-wavelength, plasmonic "resonant cavity" of the laser, allowing for strongly enhanced light-matter interaction and sub-wavelength size. We have shown that both the geometrical twist-angle of the crosses wire meta-atoms as well as the polarization of an incident pump radiation can be used to control the emission polarization state of the laser from linear to circular and to switch from right- to left-circular polarized output. Furthermore, we have shown purely circularly polarized lasering output for both pulsed and continuous-wave pump operation. We believe that our findings could guide new experimental efforts towards realization of polarization controllable circularly polarized output ultra thin surface lasers using resonant chiral metasurfaces, which offer access to a plethora of exciting photonic applications.

This research work was supported by the Department of Energy (Basic Energy Sciences, Division of Materials Sciences and Engineering) under Contact No DE-AC02-07CH11358. Work at FORTH was supported by the  European Union project FABulous (HORIZON-CL4-2022-TWIN-TRANSITION-01-02, GA:101091644), the Horizon 2020 RISE Project (CHARTIST,
101007896) as well as by the Hellenic Foundation for Research and Innovation (H.F.R.I.) under the "Basic Research Financing", National Recovery and Resilience Plan (Greece 2.0), (Project Number: 14830, PhoToCon). We dedicate this manuscript to the memory of Prof. Costas Soukoulis,
a revered leader, colleague, teacher, and friend,
whose contributions greatly enriched our collaborative works.

%\textit{Physical Review} 

\nocite{*}

\bibliography{apssamp}% Produces the bibliography via BibTeX.

\end{document}